\documentclass[preprint,12pt]{elsarticle}

\usepackage{amssymb}
\usepackage{xcolor}
\newcommand{\eref}[1]{Eq.~(\ref{#1})}
\newcommand{\fref}[1]{Fig.~(\ref{#1})}

\usepackage{amsmath}

\journal{Journal of Subatomic Particles and Cosmology}
\begin{document}

\begin{frontmatter}

\title{Finite Temperature Quarkonia Spectral Functions in the Pseudoscalar Channel}

\author[a]{Dibyendu Bala}
\author[a,b]{Sajid Ali}
\author[a]{Olaf Kaczmarek}
\author[a]{Pavan}

\address[a]{Bielefeld University, Faculty of Physics, Bielefeld, 33615, North Rhine-Westphalia, Germany}
\address[b]{Government College University Lahore, Department of Physics, Lahore 54000, Pakistan}
\author[]{\vspace{0.1cm}\\ \textbf{HotQCD Collaboration}} 

\begin{abstract}
Quarkonia, the bound states of heavy quark-antiquark pairs, are important tools for studying the quark-gluon plasma (QGP). In this study, we examine the behavior of in-medium quarkonium bound states in the QGP by analyzing their spectral functions at two temperatures, $T = 220\,\textrm{MeV}$ and $T = 293\,\textrm{MeV}$. We use physics-motivated information to reconstruct the spectral function from the Euclidean lattice correlator. Near the threshold, the spectral function is estimated through a complex potential, determined non-perturbatively from Wilson line correlators. Our results show that the real part of the potential undergoes color screening above $T_{pc}$, while the imaginary part grows rapidly with increasing distance and temperature. For the ultraviolet (UV) part of the spectral function, we use the perturbative vacuum spectral function, as the temperature effects are suppressed in this region. In the absence of a transport peak in the pseudoscalar channel, we find that this combination effectively describes the pseudoscalar correlator on the lattice, calculated using relativistic quark fields. Our results show that pseudoscalar charmonium ($\eta_c$) experiences significant thermal effects, as indicated by the broadening of the $\eta_c(1S)$ state. In contrast, the $\eta_b(1S)$ state remains intact, with a sharp bound state peak.
\end{abstract}

\begin{keyword}
Lattice QCD, Quarkonia, Non-perturbative potential, Color Screening, Spectral function
\end{keyword}

\end{frontmatter}

\section{Introduction}
\label{sec1}
Among various probes, quarkonia have played a crucial role in understanding the quark-gluon plasma produced in heavy-ion collision experiments at RHIC \cite{STAR:2019fge} and LHC \cite{CMS:2018zza}. Quarkonia are formed during the early stages of heavy-ion collisions through hard scattering processes and subsequently propagate through the quark-gluon plasma. Interactions with the plasma suppress the net yield of quarkonia observed in heavy-ion collisions compared to expectations from proton-proton collisions. This suppression is considered a key signal of plasma formation, as the quark-gluon plasma induces color screening, inhibiting quarkonium binding \cite{Matsui:1986dk}. However, this concept of dissociation has been refined to account additional effects such  as Landau damping \cite{Laine:2006ns} caused by inelastic scattering of spatial gluons and singlet-to-octet transition \cite{Brambilla:2008cx}, which contributes to the thermal decay of quarkonia within the plasma. 

To understand this suppression, one needs to know the real-time dynamics, which are encoded in the spectral function of quarkonia. In this proceeding, we present results for quarkonia spectral functions obtained from lattice QCD. On the lattice, one computes the correlation function in imaginary time, which is related to the spectral function by the following relation:

\begin{equation}
C(\tau)=\int_{0}^{\infty} \frac{d\omega}{\pi} \rho(\omega) \frac{\cosh[\omega(\tau-\frac{1}{2T})]}{\sinh[\frac{\omega}{2T}]}.
\end{equation}

Extracting the spectral function from lattice QCD using the above relation is an ill-posed problem, as many spectral functions can reproduce the same lattice correlator within error bars. Consequently, additional physical input is necessary for spectral reconstruction. Significant progress has been made in this area through Bayesian analysis, such as the Maximum Entropy Method (MEM) \cite{Asakawa:2003re,Datta:2003ww} and Bayesian Reconstruction (BR) technique \cite{Kim:2014iga}. 

In this work, we follow the approach outlined in \cite{Burnier:2017bod,Ali:2023kmr} and calculate the spectral function by combining contributions from different energy regions. The spectral functions at high energy $(\omega \gg 2 M_q)$ can be calculated using vacuum perturbation theory as thermal effects are suppressed in this region. Near $\omega \sim 2\,M_q$, naive perturbation theory breaks down and the spectral function can be calculated by solving a Schrödinger equation with a thermal potential, which is essentially resumming ladder like gluon exchange diagrams between the quark-antiquark pair \cite{Burnier:2007qm, Mocsy:2007yj}.

The $\omega \ll 2 M_q$ region however depends on the spectral function channel. For example, in the vector channel, the  $\omega \sim 0$ region involves transport contributions, where the width is related to the heavy quark diffusion coefficient, which has been extensively studied on the lattice \cite{Banerjee:2011ra, Banerjee:2022gen, Banerjee:2022uge,Altenkort:2020fgs,Altenkort:2023oms, Altenkort:2023eav, Brambilla:2022xbd, Pandey:2023dzz}. In contrast, the pseudoscalar channel does not have transport contributions, making it more suitable for studying bound states near the threshold.

Here we will calculate the spectral function near the threshold region using a non-perturbative potential.  However, calculating this potential also poses challenges, as it requires extracting the spectral function of the Wilson loop \cite{Rothkopf:2011db}. Similar to quarkonia correlator there are studies for the extraction of thermal potential using MEM and BR techniques \cite{Rothkopf:2011db, Burnier:2014ssa}. In this work, we adopt the method proposed in \cite{Bala:2019cqu}, which is guided by physical insights and has also been used to calculate the potential in the color-octet channel \cite{Bala:2020tdt}.

For the calculation of the correlation functions we use gauge field configurations generated with the HISQ action with (2+1)-flavors with a physical strange quarks mass and a pion mass of 320 MeV, with a lattice spacing of \( a = 0.028 \, \text{fm} \). 
The pseudo-critical temperature is around 180 MeV for these parameters. These configurations have already been used in other papers by the HotQCD collaboration \cite{Altenkort:2023oms}, \cite{Bazavov:2023dci}, \cite{Ali:2024xae}.

On these lattices, quarkonium correlators are calculated using clover-improved Wilson fermions, where we use a tadpole-improved clover coefficient, \( c_{SW} \). The hopping term coefficient, \( \kappa \), was tuned to reproduce the experimental spin-averaged value of the quarkonium $1S$ states. The temperatures for which we present results are 220 MeV and 293 MeV, corresponding to temporal extents of \( N_\tau = 32 \) and \( N_\tau = 24 \), respectively. The Wilson line correlators have also been measured on these configurations for the extraction of the thermal potential. Some measurements of the Wilson line correlators used in this paper, have been taken from \cite{Bazavov:2023dci}.
 
This contribution is organized as follows: In the next section, we outline the method for calculating the thermal potential from lattice and show our results on potential. In section \ref{spfn}, we use this potential to reconstruct the spectral function for the pseudoscalar channel and compare the correlators from this spectral function with those calculated directly on the lattice.

\section{Thermal potential}
\label{subsec1}

\begin{figure*}
    \centering
    \includegraphics[width=6.5cm]{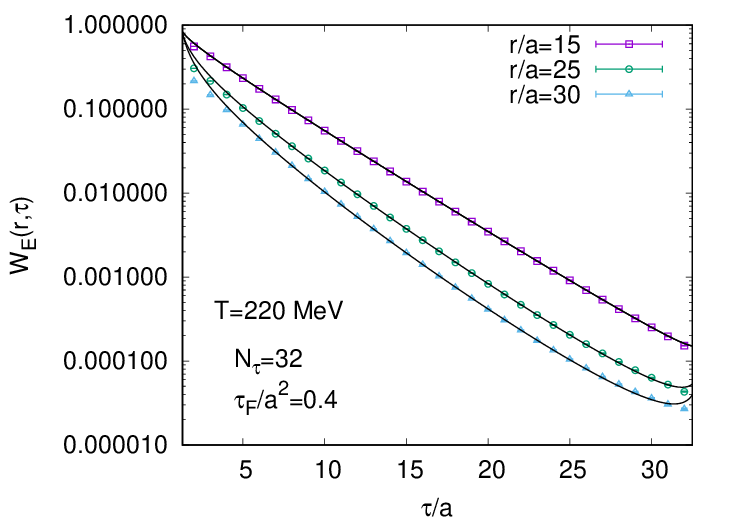}
    \includegraphics[width=6.5cm]{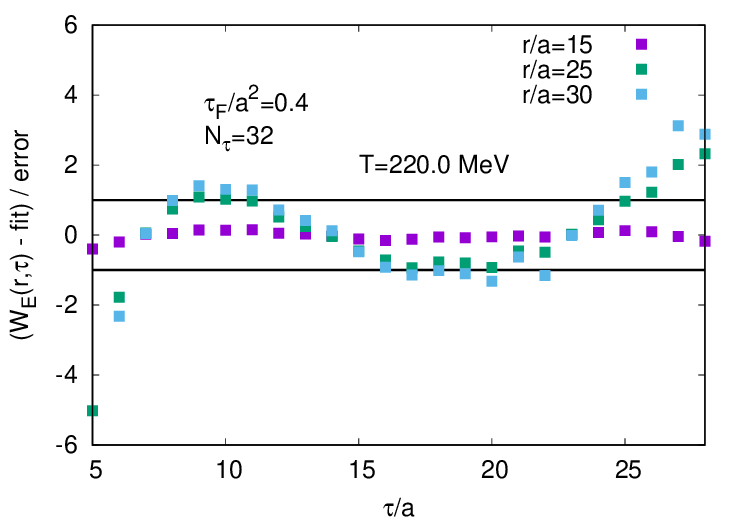}
	\caption{(Left) Fitting of Wilson line correlator with the ansatz in \eref{wparam} at $T=220 \text{MeV}$ at few distances (Right) Relative error of the fit is shown as a function of $\tau$.
	}
\label{fit}
\end{figure*}

In this section, we outline the calculation of the thermal potential as described in \cite{Bala:2019cqu}. The non-perturbative thermal potential is defined as the long-time derivative of the Wilson loop in real time \cite{Laine:2006ns}:

\begin{equation}
    V(r) = i \lim_{t \to \infty} \frac{\partial \log[W_{M}(r, t)]}{\partial t}=V_{re}(r)-i V_{im}(r).
\label{limit}
\end{equation}

The imaginary-time Wilson loop calculated on the lattice is related to the real-time Wilson loop through the following relations \cite{Rothkopf:2011db}:

\begin{align}
W_{E}(r, \tau) &= \int_{-\infty}^{\infty} d\omega \, \rho_{W}(r, \omega) \exp(-\omega \tau), \\
W_{M}(r, t) &= \int_{-\infty}^{\infty} d\omega \, \rho_{W}(r, \omega) \exp(-i \omega t).
\label{wsp}
\end{align}

An important condition on the spectral function $\rho_{W}$ is the existence of the limit in \eref{limit}. Hard-Thermal-Loop (HTL) perturbative calculations of the Euclidean Wilson loop, and the corresponding analytic continuation $\tau \to it$, show the existence of this limit \cite{Laine:2006ns}. 
Motivated by this observation, one can derive the following parametrization \cite{Bala:2019cqu}:

\begin{equation}
W(r, \tau) = A \exp\left(-V_{\text{re}}(r) \tau + \frac{V_{\text{im}}(r)}{\pi\,T} \log(\sin\left(\pi\,\tau\,T\right))+...\right).
\label{wparam}
\end{equation}

It is evident from this form that it leads to a well-defined potential. Here, we used the Wilson line correlator fixed in the Coulomb gauge, rather than the Wilson loop, to avoid significant UV contamination \cite{Burnier:2013fca}. In addition we have also used gradient flow to improve the signal for Wilson line correlator at large separation.

We fitted the Wilson line correlator using the above parameterization, and the resulting fit is shown in \fref{fit}, for $T = 220~\text{MeV}$ ($1.2 T_c$). The $\chi^2/\text{dof}$ for all these fits are around 1.

\begin{figure*}
    \centering
    \includegraphics[width=6.5cm]{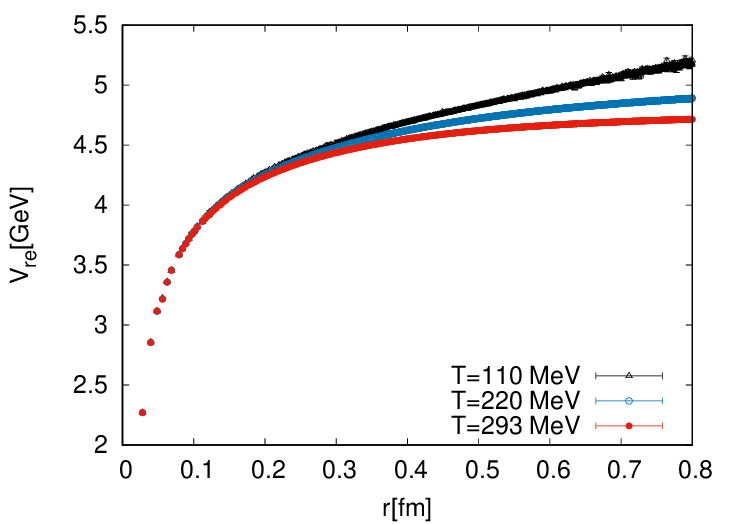}
    \includegraphics[width=6.5cm]{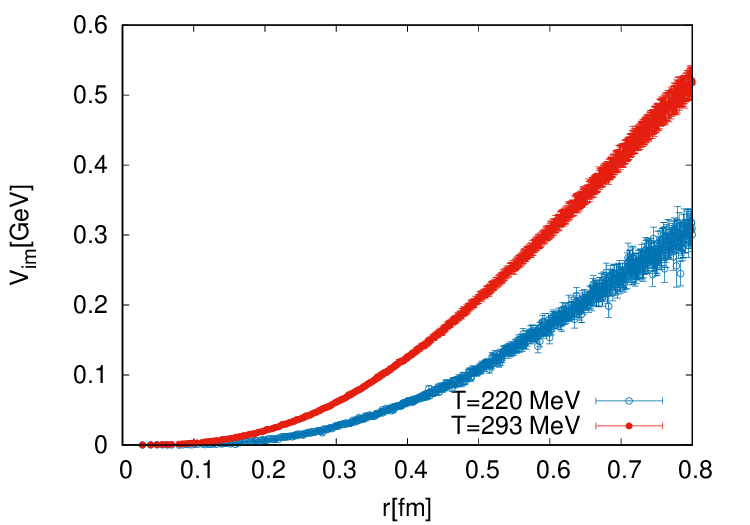}
	\caption{The potential obtained from the parametrization of $\eref{wparam}$. (Left) The real part plotted for various temperature shows color screening above the crossover temperature ($180 \text{MeV}$). (Right) The imaginary part increases with both temperature and distance.
	}
\label{potential}
\end{figure*}
The fit range used in this figure is from $\tau/a = 6$ to $26$, as the parameterization is only valid in the region $0\ll\tau\ll \frac{1}{T}$. We also verified that changing the fit range does not affect the potential, indicating that the fit is stable. 
The gradient flow distorts the short-distance behavior of the real part of the potential. As a result, we replace only the large-distance part of the potential with the gradient-flowed potential, which effectively reduces the error at large distances. To obtain the real part therefore we smoothly match the zero-flow potential at short distances with the gradient-flow potential at larger distances. For the imaginary part, we do not observe significant flow-time dependence within the error bars. Nevertheless, the final imaginary part is obtained by performing a zero-flow-time extrapolation, using a linear fit in flow time. The resulting potentials obtained after this is shown in \fref{potential}. On the left panel of \fref{potential}, we see that the real part of of the potential shows color screening with increasing temperature. However the strength of screening of the potential is different compared to perturbative case. The screening of the potential is consistent with the observations in \cite{Burnier:2014ssa, Bala:2019cqu, Larsen:2024wgw}, but contradicts \cite{Bazavov:2023dci}, where no screening is observed up to $T \sim 350 \, \text{MeV}$. This contradiction illustrates the ill-posed nature of the inverse problem in \eref{wsp}. It also highlights the crucial role of incorporating physics-driven input to obtain meaningful results when performing the analytic continuation of Euclidean-time lattice correlators to real-time correlators.

The imaginary part of the potential increases with temperature and distance. The imaginary part turns out to be much larger compared to the perturbative value at this temperature indicating larger damping rate of quarkonium bound states.
 \section{Pseudo-Scalar Quarkonium Spectral functions}
 \label{spfn}
In this section we show the calculation of the pseudo-scalar spectral function for the lattice correlation function for bottomonium and charmonium defined as,  
\begin{equation}
C_{\text{PS}}(\tau) = M_{B}^2 \int d^3 \vec{x} \, 
\langle \bar{\psi}(\vec{x}, \tau) \gamma_5 \psi(\vec{x}, \tau) 
\bar{\psi}(\vec{0}, 0) \gamma_5 \psi(\vec{0}, 0) \rangle_T \, .
\label{pscorr}
\end{equation}
Here $M_{B}$ is bare quark mass, that ensures the pseudoscalar current becomes divergence-free after coupling renormalization \cite{Burnier:2017bod}. $\psi$ is either bottom quark or charm quark field. 

As mentioned in the introduction the spectral function near $\omega \sim 2 M_q$ region can be obtained by solving the Schrödinger equation with thermal potential given in \fref{potential} as follows \cite{Burnier:2007qm},
\begin{equation}
2\, M_q-\frac{\nabla^2}{M_q} C_{>}(r, t) + V(r) C_{>}(r, t) = 
i \frac{\partial C_{>}(r, t)}{\partial t},
\,
C_{>}(\vec{r}, 0) = -2\,M_{q}^2\,N_c\,\delta^3(\vec r).
\end{equation}
Here $C_{>}(r,t)$ is the thermal average real time forward correlation function of the gauge invariant point-spilt version of pseudoscalar current. The pseuoscalar channel spectral function near $\omega \sim 2\,M_q$ is then related to the Fourier transform of the the function $C_{>}(r,t)$:
\begin{equation}
\rho_{\text{PS}}(\omega)=\lim_{r\to 0}\int_{-\infty}^{\infty}\,d\omega\, C_{>}(r,t) \exp(i\omega t)
\end{equation}

Before using the potential, lattice artifacts at short distances need to be removed and the additive renormalization of the lattice potential must be fixed also. Additionally, the pole mass to be used in the Schrodinger equation needs to be determined .  

To eliminate lattice artifacts at short distances, the temperature independent part of the potential is replaced by matching it at around $r\sim 0.1 \,\text{fm}$, to the three-loop renormalon-subtracted (RS) perturbative potential from \cite{Smirnov:2009fh}. However, this matching still leaves an uncertainty in the additive renormalization of $O(\Lambda_{\text{QCD}})$, originating from the renormalon pole \cite{Sumino:2014qpa}.  

To determine the additive constant, we used the masses of $\eta_b(1S)$ and $\eta_c(1S)$ obtained from the pseudoscalar correlation function measured on the lattice at $T = 110\,\text{MeV}$. These masses are $M_{\eta_b} = 9.4 \pm 0.02$~GeV and $M_{\eta_c} = 2.97 \pm 0.02$~GeV. These values are consistent with the PDG values within error bars. We set the bottom quark mass to $m_b = 4.78$~GeV and solved the Schrödinger equation using the zero temperature potential. The additive constant was tuned such that $\eta_b(1S)$ mass agree the lattice-determined mass of $\eta_{b}(1S)$ state. 

\begin{figure*}
    \centering
    \includegraphics[width=6.5cm]{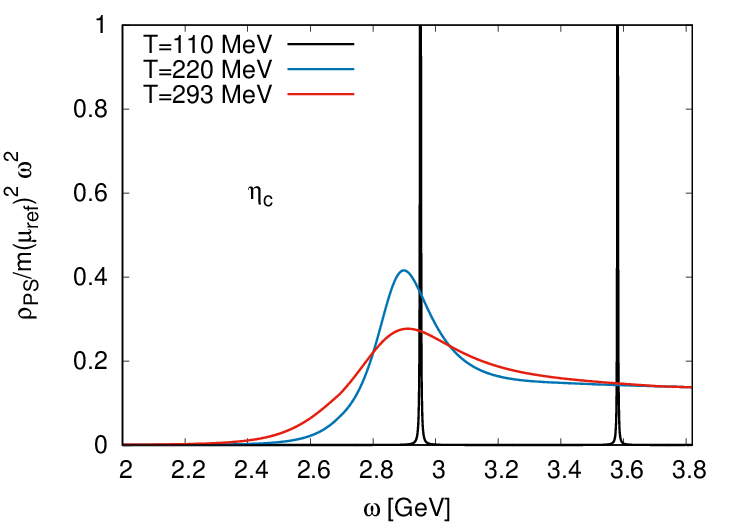}
    \includegraphics[width=6.5cm]{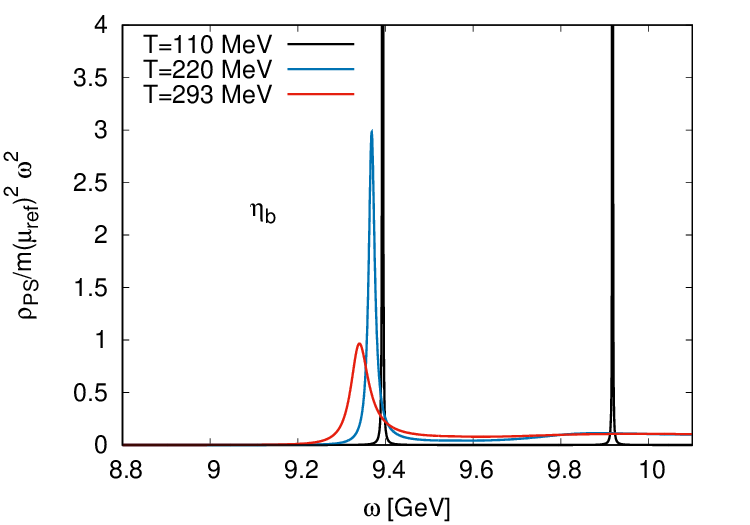}
	\caption{ The spectral functions of $\eta_c$ (left) and $\eta_b$ (right) are shown below and above the pseudo-critical temperature $T_{pc} = 180\, \text{MeV}$. Here, $m(\mu_{\text{ref}})$ represents the $\overline{\text{MS}}$ mass calculated at the reference scale $\mu_{\text{ref}} = 2 \, \text{GeV}$. This running mass is used instead of the pole mass to ensure the perturbative vacuum spectral function is well-converged at high $\omega$.
	}
\label{sp_temp}
\end{figure*}
Once the additive constant is fixed, the remaining charm quark mass was obtained again by solving the Schrödinger equation, such that the zero-temperature $\eta_c(1S)$ mass was reproduced. This correspond to a charm mass $m_c = 1.34(2)\,\text{GeV}$. This fixes all the quark masses and the additive constant.
Using these masses and the additive constant, we have then solved the Schrödinger equation using the finite temperature potential of \fref{potential} to get the spectral function in the threshold region. Near $\omega \ll 2\,M_q$ the spectral function is exponentially suppressed, however Schrodinger description overestimate the spectral function in this region as Schrodinger description is not a valid description in this region. To model this suppression the imaginary part potential is multiplied by the factor $\exp(\frac{2\,M_q-\omega}{T})$ in the region $\omega < 2\,M_q$ region. The UV part of the spectral function has been taken from vacuum perturbation theory and has been matched with the spectral function from thermal potential according to the following form,
\begin{equation}
\rho_{matched}=A_{0}\,\rho_{T}(\omega) \theta(\omega_{match}-\omega)+\rho_{vac}(\omega)\theta(\omega-\omega_{match}),
\end{equation}
 where $\omega_{match} \sim 2.6(1) M$. The final spectral function is shown in \fref{sp_temp}, with the left panel for the charm quark and right panel for bottom quark. In these figures, we have also plotted the spectral function at $110\,\text{MeV}$ (below $T_{pc}$), which shows sharp peaks at the bound state position. Above the crossover temperature, we observe that the excited state peak has disappeared, and the sharp ground state peak has broadened, with a shift in the peak position towards a lower value. The broadening is much more pronounced for the $\eta_c(1S)$ state than for the $\eta_b(1S)$ state, as expected, due to the much smaller charm quark mass compared to the bottom quark mass. Additionally, as the temperature increases, the thermal width grows. At $T = 293\,\text{MeV}$, the thermal width of the $\eta_c(1S)$ state becomes very large, indicating that this state is already close to its melting temperature, whereas a still sharp bound state peak is observed for the $\eta_b(1S)$ state.     

\section{Comparing with lattice correlator}
\begin{figure*}
    \centering
    \includegraphics[width=6.5cm]{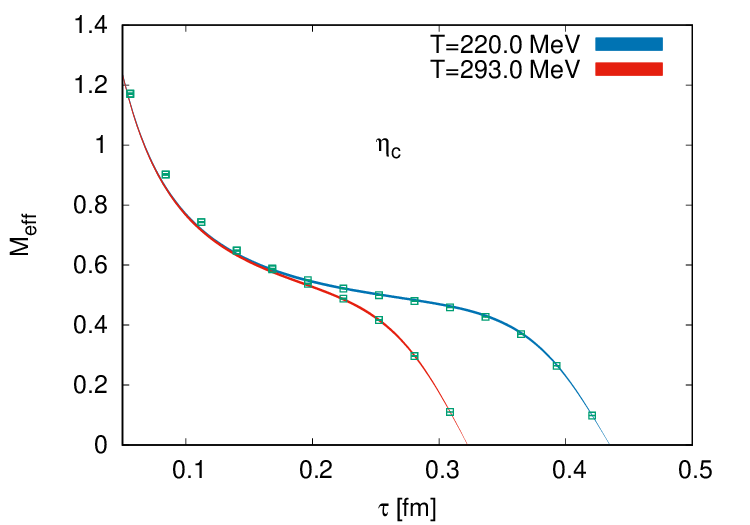}
    \includegraphics[width=6.5cm]{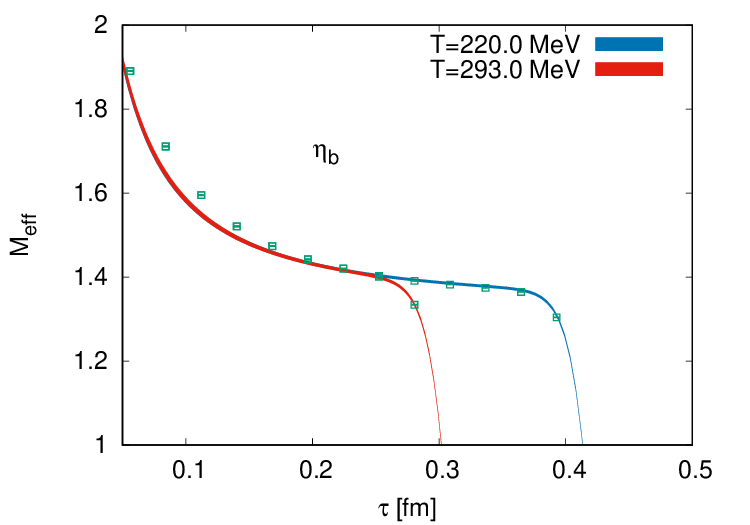}
	\caption{ Comparison of the effective masses obtained from the spectral function in \fref{sp_temp} with the corresponding effective masses calculated from the lattice correlator, as defined in \eref{pscorr}. On the left (right) panel we show $\eta_c$ ($\eta_b$) correlator. We find reasonably good agreement after $\tau \sim 0.22 \, \text{fm}$. 	The short distance points are not expected to reproduce, as the lattice correlator is not continuum-extrapolated. }
\label{comp}
\end{figure*}
Now we compare the correlator obtained from the spectral functions in \fref{sp_temp} with the corresponding correlator calculated on the lattice. On the lattice, however, we calculate only the connected part of the correlation function. The disconnected part of the correlation, however, is expected to be small in the heavy quark region.  Since lattice correlator has finite multiplicative renormalization factor, we do the comparison in the effective mass level defined as,
\begin{equation}
M_{\text{eff}}(\tau)=\log\left(\frac{C^{PS}(\tau)}{C^{PS}(\tau+a)}\right)
\end{equation}
In \fref{comp} we show these comparisons for charm and bottom quark. One can see that these spectral functions are indeed able to predict the correlation functions. The error band in the prediction is mostly due to error from the zero temperature $1S$ masses.  A quantitative measure of the prediction , 
\begin{equation}
\chi^2 = \sum_{\tau > 0.224 \text{fm}} \frac{(\text{data} - \text{prediction})^2}{(\text{err}_{\text{data}} + \text{err}_{\text{prediction}})^2},
\end{equation}
yields $\chi^2/\text{dof} = 0.3- 0.9$. The short $\tau$ part is not expected to be described by the spectral functions \fref{sp_temp}, because of the cut-off effect present in this region.

This consistency at large $\tau$ shows that the non-perturbative screened potential is a valid thermal potential that is consisted with lattice QCD correlator defined in \eref{pscorr} . 

\section{Summary}
In this proceeding, we present new results on quarkonia spectral functions in the pseudoscalar channel. We calculate the spectral function using the thermal potential obtained from the Wilson line correlator. Our parametrization of Wilson line correlator as given in \eref{wparam}, fits the correlator over a reasonably large $\tau$ window as shown in \fref{fit}. We find that the real part of the potential supports color screening once the temperature rises above the chiral crossover temperature. The imaginary part of the potential exhibits a rapid increase with both distance and temperature.

The spectral function obtained from the thermal potential, valid near the threshold, is matched with the ultraviolet perturbative vacuum spectral function. The resulting spectral function is shown in \fref{sp_temp}, which shows significant thermal modifications to the $\eta_c(1S)$ state compared to the $\eta_b(1S)$ state. At our highest temperature of $T = 293 \, \mathrm{MeV}$, the $\eta_b(1S)$ state still displays a reasonably sharp bound state peak, while the $\eta_c(1S)$ state is already near its melting point. Using the spectral function, we compare our results directly with the corresponding lattice correlator. The effective mass comparison, shown in \fref{comp}, indicates consistency between the spectral function and the lattice correlator in the large-$\tau$ region. The small-$\tau$ region does not match due to lattice artifacts, as expected. This comparison demonstrates that the color screening observed in the real part of the potential (see \fref{potential}) is consistent with lattice correlator defined in \eref{pscorr}.

Further technical details will be provided in a follow-up paper.

\section{Acknowledgments}
The gauge configuration generation and Wilson line correlator measurements were performed using SIMULATeQCD~\cite{HotQCD:2023ghu}. These configurations has also been used in other HotQCD studies \cite{Altenkort:2023oms}, \cite{Bazavov:2023dci}, \cite{Ali:2024xae}. Some of the Wilson line correlator measurements used in this proceeding were taken from \cite{Bazavov:2023dci}, and we thank the authors of \cite{Bazavov:2023dci} for providing them. We would like to thank Luis Altenkort and Hai-Tao Shu for their work on integrating the mesonic correlator into the QUDA \cite{Clark:2009wm} code, which has been used for the measurement of charm and bottom-point correlation functions. D.~B. would like to thank Saumen Datta for various discussions. The authors 
acknowledge support by the Deutsche Forschungsgemeinschaft (DFG, German Research Foundation) through the CRC-TR 211 `Strong-interaction matter under extreme conditions' – project number 315477589 – TRR 211.
For the computational work we used the Bielefeld GPU cluster and the LUMI-G supercomputer. We acknowledge the EuroHPC Joint Undertaking forawarding  this  project  access  to  the  EuroHPC  supercomputer  LUMI-G,hosted by CSC (Finland) and the LUMI consortium through a EuroHPCExtreme Scale Access call.


\begin{thebibliography}{00}


\bibitem{STAR:2019fge}
J.~Adam \textit{et al.} [STAR],
Phys. Lett. B \textbf{797} (2019), 134917
doi:10.1016/j.physletb.2019.134917
[arXiv:1905.13669 [nucl-ex]].
\bibitem{CMS:2018zza}
A.~M.~Sirunyan \textit{et al.} [CMS],
Phys. Lett. B \textbf{790} (2019), 270-293
doi:10.1016/j.physletb.2019.01.006
[arXiv:1805.09215 [hep-ex]].
\bibitem{Matsui:1986dk}
T.~Matsui and H.~Satz,
Phys. Lett. B \textbf{178} (1986), 416-422
doi:10.1016/0370-2693(86)91404-8
\bibitem{Laine:2006ns}
M.~Laine, O.~Philipsen, P.~Romatschke and M.~Tassler,
JHEP \textbf{03} (2007), 054
doi:10.1088/1126-6708/2007/03/054
[arXiv:hep-ph/0611300 [hep-ph]].
\bibitem{Brambilla:2008cx}
N.~Brambilla, J.~Ghiglieri, A.~Vairo and P.~Petreczky,
Phys. Rev. D \textbf{78} (2008), 014017
doi:10.1103/PhysRevD.78.014017
[arXiv:0804.0993 [hep-ph]].
\bibitem{Asakawa:2003re}
M.~Asakawa and T.~Hatsuda,
Phys. Rev. Lett. \textbf{92} (2004), 012001
doi:10.1103/PhysRevLett.92.012001
[arXiv:hep-lat/0308034 [hep-lat]]
\bibitem{Datta:2003ww}
S.~Datta, F.~Karsch, P.~Petreczky and I.~Wetzorke,
Phys. Rev. D \textbf{69} (2004), 094507
doi:10.1103/PhysRevD.69.094507
[arXiv:hep-lat/0312037 [hep-lat]].
\bibitem{Kim:2014iga}
S.~Kim, P.~Petreczky and A.~Rothkopf,
Phys. Rev. D \textbf{91} (2015), 054511
doi:10.1103/PhysRevD.91.054511
[arXiv:1409.3630 [hep-lat]].
\bibitem{Burnier:2017bod}
Y.~Burnier, H.~T.~Ding, O.~Kaczmarek, A.~L.~Kruse, M.~Laine, H.~Ohno and H.~Sandmeyer,
JHEP \textbf{11} (2017), 206
doi:10.1007/JHEP11(2017)206
[arXiv:1709.07612 [hep-lat]].
\bibitem{Ali:2023kmr}
S.~Ali \textit{et al.} [HotQCD],
Few Body Syst. \textbf{64} (2023) no.3, 52
doi:10.1007/s00601-023-01833-w
[arXiv:2305.06907 [hep-lat]].
\bibitem{Burnier:2007qm}
Y.~Burnier, M.~Laine and M.~Vepsalainen,
JHEP \textbf{01} (2008), 043
doi:10.1088/1126-6708/2008/01/043
[arXiv:0711.1743 [hep-ph]].
\bibitem{Banerjee:2011ra}
D.~Banerjee, S.~Datta, R.~Gavai and P.~Majumdar,
Phys. Rev. D \textbf{85} (2012), 014510
doi:10.1103/PhysRevD.85.014510
[arXiv:1109.5738 [hep-lat]].
\bibitem{Banerjee:2022gen}
D.~Banerjee, R.~Gavai, S.~Datta and P.~Majumdar,
doi:10.1016/j.nuclphysa.2023.122721
[arXiv:2206.15471 [hep-ph]].
\bibitem{Banerjee:2022uge}
D.~Banerjee, S.~Datta and M.~Laine,
JHEP \textbf{08} (2022), 128
doi:10.1007/JHEP08(2022)128
[arXiv:2204.14075 [hep-lat]].
\bibitem{Altenkort:2020fgs}
L.~Altenkort, A.~M.~Eller, O.~Kaczmarek, L.~Mazur, G.~D.~Moore and H.~T.~Shu,
Phys. Rev. D \textbf{103} (2021) no.1, 014511
doi:10.1103/PhysRevD.103.014511
[arXiv:2009.13553 [hep-lat]].
\bibitem{Altenkort:2023oms}
L.~Altenkort \textit{et al.} [HotQCD],
Phys. Rev. Lett. \textbf{130} (2023) no.23, 231902
doi:10.1103/PhysRevLett.130.231902
[arXiv:2302.08501 [hep-lat]].
\bibitem{Altenkort:2023eav}
L.~Altenkort \textit{et al.} [HotQCD],
Phys. Rev. Lett. \textbf{132} (2024) no.5, 051902
doi:10.1103/PhysRevLett.132.051902
[arXiv:2311.01525 [hep-lat]].
\bibitem{Brambilla:2022xbd}
N.~Brambilla \textit{et al.} [TUMQCD],
Phys. Rev. D \textbf{107} (2023) no.5, 054508
doi:10.1103/PhysRevD.107.054508
[arXiv:2206.02861 [hep-lat]].

\bibitem{Pandey:2023dzz}
H.~Pandey, S.~Schlichting and S.~Sharma,
Phys. Rev. Lett. \textbf{132} (2024) no.22, 222301
doi:10.1103/PhysRevLett.132.222301
[arXiv:2312.12280 [hep-lat]].

\bibitem{Rothkopf:2011db}
A.~Rothkopf, T.~Hatsuda and S.~Sasaki,
Phys. Rev. Lett. \textbf{108} (2012), 162001
doi:10.1103/PhysRevLett.108.162001
[arXiv:1108.1579 [hep-lat]].
\bibitem{Burnier:2014ssa}
Y.~Burnier, O.~Kaczmarek and A.~Rothkopf,
Phys. Rev. Lett. \textbf{114} (2015) no.8, 082001
doi:10.1103/PhysRevLett.114.082001
[arXiv:1410.2546 [hep-lat]].
\bibitem{Bala:2019cqu}
D.~Bala and S.~Datta,
Phys. Rev. D \textbf{101} (2020) no.3, 034507
doi:10.1103/PhysRevD.101.034507
[arXiv:1909.10548 [hep-lat]].

\bibitem{Bala:2020tdt}
D.~Bala and S.~Datta,
Phys. Rev. D \textbf{103} (2021) no.1, 014512
doi:10.1103/PhysRevD.103.014512
[arXiv:2009.00773 [hep-lat]].

\bibitem{Burnier:2013fca}
Y.~Burnier and A.~Rothkopf,
Phys. Rev. D \textbf{87} (2013), 114019
doi:10.1103/PhysRevD.87.114019
[arXiv:1304.4154 [hep-ph]].

\bibitem{Smirnov:2009fh}
A.~V.~Smirnov, V.~A.~Smirnov and M.~Steinhauser,
Phys. Rev. Lett. \textbf{104} (2010), 112002
doi:10.1103/PhysRevLett.104.112002
[arXiv:0911.4742 [hep-ph]].
\bibitem{Sumino:2014qpa}
Y.~Sumino,
[arXiv:1411.7853 [hep-ph]].
\bibitem{HotQCD:2023ghu}
L.~Mazur \textit{et al.} [HotQCD],
Comput. Phys. Commun. \textbf{300} (2024), 109164
doi:10.1016/j.cpc.2024.109164
[arXiv:2306.01098 [hep-lat]].
\bibitem{Bazavov:2023dci}
A.~Bazavov \textit{et al.} [HotQCD],
Phys. Rev. D \textbf{109}, no.7, 074504 (2024)
doi:10.1103/PhysRevD.109.074504
[arXiv:2308.16587 [hep-lat]].
\bibitem{Ali:2024xae}
S.~Ali \textit{et al.} [HotQCD],
Phys. Rev. D \textbf{110}, no.5, 5 (2024)
doi:10.1103/PhysRevD.110.054518
[arXiv:2403.11647 [hep-lat]].
\bibitem{Mocsy:2007yj}
A.~Mocsy and P.~Petreczky,
Phys. Rev. D \textbf{77}, 014501 (2008)
doi:10.1103/PhysRevD.77.014501
[arXiv:0705.2559 [hep-ph]].
\bibitem{Larsen:2024wgw}
R.~N.~Larsen, G.~Parkar, A.~Rothkopf and J.~H.~Weber,
Phys. Rev. D \textbf{110} (2024) no.11, 114501
doi:10.1103/PhysRevD.110.114501
[arXiv:2402.10819 [hep-lat]].
\bibitem{Clark:2009wm}
M.~A.~Clark \textit{et al.} [QUDA],
Comput. Phys. Commun. \textbf{181} (2010), 1517-1528
doi:10.1016/j.cpc.2010.05.002
[arXiv:0911.3191 [hep-lat]].
\end{thebibliography}
\end{document}